\documentclass[3p,times]{elsarticle}

\usepackage{ecrc}

\volume{00}

\firstpage{1}

\journalname{Physics Procedia}

\runauth{B.A.\ Berg}

\jid{phpro}

\jnltitlelogo{Physics Procedia}

\usepackage{amssymb}

\usepackage[figuresright]{rotating}

\begin{document}

\begin{frontmatter}



\title{Display of probability densities for data from a continuous 
distribution}


\author[1]{Bernd A.\ Berg}

\address[1]{Department of Physics, Florida State University, 
           Tallahassee, FL 32306, USA}

\begin{abstract}
Based on cumulative distribution functions, Fourier series expansion
and Kolmogorov tests, we present a simple method to display probability
densities for data drawn from a continuous distribution. It is often
more efficient than using histograms.
\end{abstract}


\end{frontmatter}



\section{Introduction} \label{sec_intro}
We address the simple problem of displaying an empirical probability 
density (PD) $f(x)$ from data for a {\it continuous} variable $x$.
Commonly this is done using histograms. This is appropriate when $x$ 
is discrete, because there is then a natural scale. But in case of a 
continuous variable $x$, one is faced with choosing binsizes. This is 
a frustrated problem: One would like to keep the binsize small for a 
high resolution, but big to suppress statistical fluctuations. Here we 
present a method \cite{BH08} to by-pass the problem. It is based on the 
cumulative distribution function (CDF)
\begin{equation} 
  F(x)\ =\ \int_{-\infty}^x dx'f(x')\ .
\end{equation}
Given a time series of $n$ real numbers (data), a parameter free 
empirical estimate (ECDF), is well-known: The step function 
$\overline{F}(x)$ defined by increasing by $1/n$ at each data point. 
This does not help directly in getting an estimate of the probability 
density, because the derivative is a sum of Dirac delta functions. 

One needs some kind of interpolation of the CDF. This is no fun, as 
one has to decide whether the interpolation of 2, 3, 4, or $k$ points 
will work best. In contrast, plotting a histogram is simple and robust, 
but not a smooth function. Our way out relies on {\it Fourier expansion
} of the ECDF $\overline{F}(x).$ This leads to the desired smooth 
approximation as long as the expansion is sufficiently short, but 
will imitate every wiggle of the data, when carried too far. Therefore, 
one needs a cut-off criterion. We base this on the {\it Kolmogorov test
}, which tells us whether the difference between the ECDF and an 
analytical approximation of the CDF is explained by chance. {\it 
Fortran code} for our procedure \cite{BH08} is available from the 
CPC Library.

\section{(Peaked) Cumulative Distribution Functions} \label{sec_CDF}
\begin{figure}[ht] \begin{center}
\includegraphics[scale=0.8]{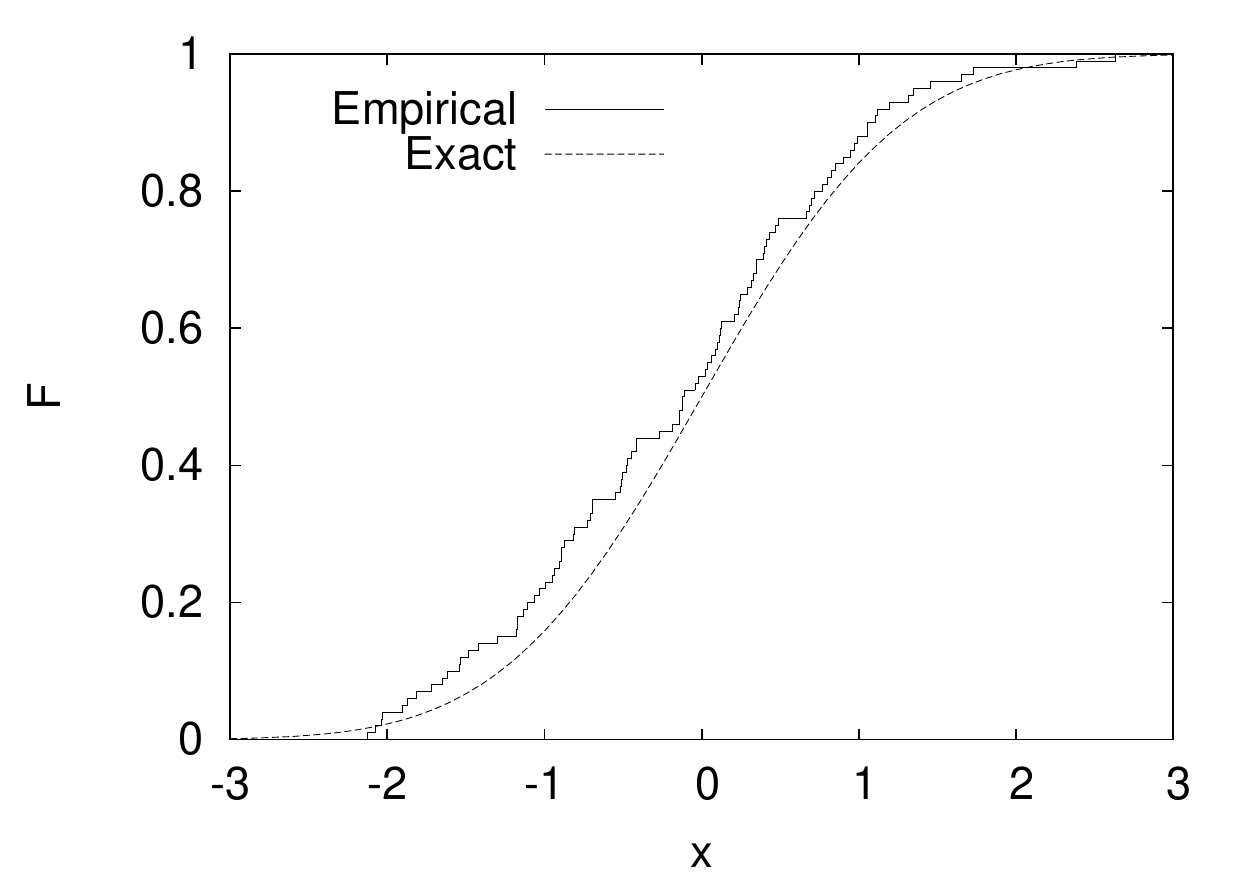}
\caption{\label{fig_GCDF100} ECDF from 100 Gaussian distributed random 
numbers together with the exact CDF.}
\end{center} \end{figure}

\begin{figure}[ht] \begin{center}
\includegraphics[scale=0.8]{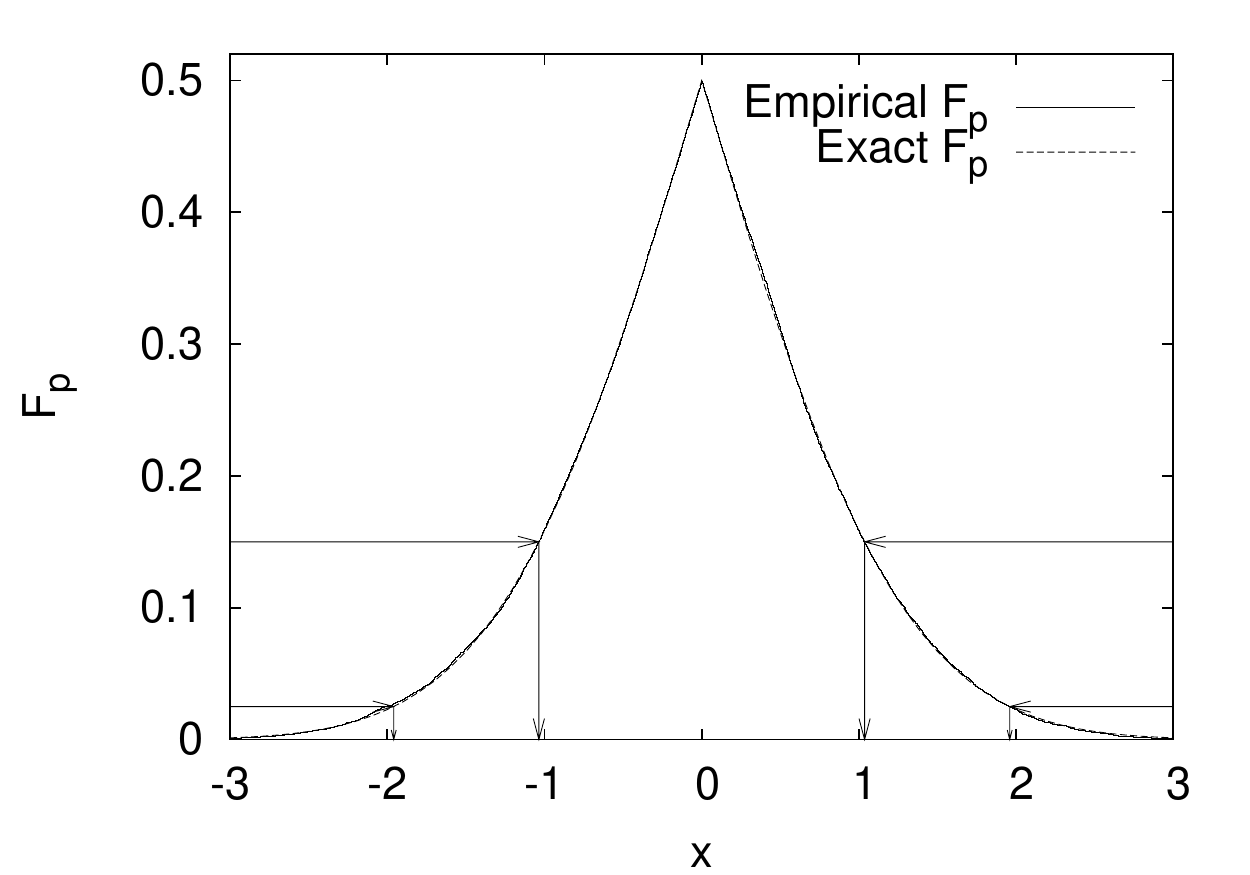}
\caption{\label{fig_GpCDF10000} Peaked ECDF from the 10,000
Gaussian random numbers versus exact Gaussian peaked CDF.
The arrows indicate 70\% and 95\% confidence intervals.}
\end{center} \end{figure}

Assume we generate $n$ random numbers $x_1$, $\cdots$, $x_n.$ We 
re-arrange the $x_i$ in increasing order ($\pi_1,\dots , \pi_n$ 
a permutation of $1,\dots ,n$):
\begin{equation} 
  x_{\pi_1} \le x_{\pi_2} \le \dots \le x_{\pi_n} 
\end{equation}
An estimator for the distribution function $F(x)$ is the ECDF
\begin{equation}
  \overline{F} (x) = {i\over n}~~~{\rm for}~~~x_{\pi_i} \le x <
  x_{\pi_{i+1}},~~~i=0, 1,\dots, n-1, n,
\end{equation}
and by definition $x_{\pi_0}=-\infty$, $x_{\pi_{n+1}}=+\infty$.
Fig.~\ref{fig_GCDF100} shows an ECDF from 100 Gaussian distributed 
random numbers generated for the probability density 
\begin{equation}
  g(x)\ =\ \frac{1}{\sqrt{2\pi}}\, \exp\left(-\frac{x^2}{2}\right)
\end{equation}
together with the exact CDF. 
The CDF is in this case determined by the error function:
\begin{equation}
  G(x) = \int_{-\infty}^x dx'g(x') = \frac{1}{2} + \frac{1}{2}\,
         {\rm erf}\left(\frac{x}{\sqrt{2}}\right)\, .
\end{equation}
The probability density of events is encoded in the slope of the ECDF. 
This makes it often difficult to read off high probability regions and, 
in particular, the median. This can be improved by switching to the 
peaked CDF \cite{B04}:
\begin{equation}
  F_p(x) = \begin{cases} F(x)\ {\rm for}\ F(x) \le {1\over 2}\,;\\
           1 - F(x)\ {\rm for}\ F(x) > {1\over 2}\,. \end{cases} 
\end{equation}
By construction the maximum of the peaked CDF is at the median
$x_{1\over 2}$~and $F_p(x_{1/2})=1/2$. Therefore, $F_p(x)$ has 
two advantages: The median is clearly exhibited and the accuracy of 
the ordinate is doubled. It looks a bit like a PD, but is in essence 
still the integrated~PD. An example from 10,000 Gaussian random numbers
is shown in Fig.~\ref{fig_GpCDF10000}.

\section{Kolmogorov Test} \label{sec_Kol}

Do empirical and exact CDFs of our two figures agree? The Kolmogorov 
test answers this question (for a review see \cite{B04}). It returns 
the {\it probability $Q$, that the difference between the analytical 
CDF and an ECDF from statistically independent data is due to chance.} 
If the analytical CDF is known and the data are sampled from this 
distribution, $Q$ is a uniformly distributed random variable in the 
range $0<Q<1$.  Turned around, if one is not sure about the exact CDF, 
or the data, or both, and $Q$ is small (say, $Q<10^{-6}$) one concludes 
that the difference between the proposed CDF and the data is presumably 
not due to chance. Kolmogorov's ingenious test relies just on the maximum
difference between the ECDF and the CDF:
\begin{equation}
  \triangle\ =\ \max_x\left|F(x)-\overline{F}(x)\right|\ .
\end{equation}
The test yields, respectively, $Q=0.19$ and $Q=0.78$ for the samples 
used in Fig.~\ref{fig_GCDF100} and~\ref{fig_GpCDF10000}. Both values
signal consistency between CDF and data.

\section{Probability Densities} \label{sec_PD}

Our method \cite{BH08} to construct an empirical probability density 
(EPD) from an ECDF consists of two steps:
\begin{enumerate}
\item Define as an initial approximation to $\overline{F}(x)$ a 
      differentiable, monotonically increasing function $F_0(x)$. 
\item Fourier expand the remainder until the Kolmogorov test yields 
      $Q\ge Q_{\rm cut}=1/2$ (there may be some flexibility in 
      lowering $Q_{\rm cut}$).
\end{enumerate}
For $F_0(x)$ we require 
\begin{eqnarray}
  F_0(x) = 0~~{\rm for}~~x\le a~~~{\rm and}~~~1~~{\rm for}~x\ge b\,,
\end{eqnarray}
where $[a,b]$ has to lie within the range of the data. For PDs 
with support on a compact interval, or with fast fall-off like for 
a Gaussian distribution, the natural choice is $a=x_{\pi_1}$ 
and $b=x_{\pi_n}$. In case of slow fall-off, like for a Cauchy
distribution, or other distributions with outliers, one has to 
restrict the analysis to $[a,b]$ regions, which are well
populated by data. 

We denote the ECDF of the range $[a,b]$ by $\overline{F}_{ab}(x)$. 
As for $F_0(x)$, by construction $\overline{F}_{ab}(x)=0$ for 
$x\le a$ and 1 for $x\ge b$. Our aim is to construct a PD estimator 
$\overline{f}_{ab}(x)$ from $\overline{F}_{ab}(x)$. In the following 
we restrict our choice of $F_0(x)$ to the straight line,
\begin{equation}
  F_0(x) = \frac{x-a}{b-a}~~{\rm for}~~a\le x\le b\,,
\end{equation}
which keeps the approach simple. More elaborate definitions will 
likely give improvements in a number of situations, but may discourage 
applications. Once $F_0(x)$ is defined, the remainder of the ECDF is 
given by
\begin{equation}
  R(x) = \overline{F}_{ab}(x)-F_0(x)\,.
\end{equation}
We expand $R(x)$ into the Fourier series
\begin{equation}
  R(x)=\sum_{i=1}^m d(i)\,\sin\left(\frac{i\,\pi\,(x-a)}{b-a}\right)\ .
\end{equation}
The cosine terms are not present due to the boundary conditions
$R(a)=R(b)=0$. The Fourier coefficients follow from
\begin{equation}
  d(i) = \sqrt{\frac{2}{b-a}} \int_a^b dx\,R(x)\,
         \sin\left(\frac{i\,\pi\,(x-a)}{b-a}\right)
\end{equation}
In our case $R(x)$ is the difference of a step function and a linear 
function. The integrals over the flat regions of the step function 
are easily calculated, and the $d(i)$ obtained by adding them up.

The Fourier expansion is useless for too large values of $m$, because 
it will then reproduce all statistical fluctuations of the data. To 
get around this problem, we perform the Kolmogorov test first between 
$\overline{F}_{ab}(x)$ and $F_0(x)$ ($m=0$), and then each time $m$ 
is incremented from $m\to m+1$.
Once $Q\ge Q_{\rm cut} = 1/2$ is reached, we know that the information 
left in the data is statistical noise and the expansion is terminated. 
The thus obtained \r smooth estimate \b of the CDF, 
\begin{equation}
  F_{\rm estimate}(x) = F_0(x)+R(x)\,, 
\end{equation}
yields $\overline{f}_{ab}(x)$ by differentiation.

We attach {\it error bars} to the estimate of the PD by dividing the 
(unsorted) original data into {\it jackknife} blocks and repeat the 
analysis for each block. Comparing the thus obtained function values, 
error bars follow in the usual jackknife way. An example for the 
Gaussian distribution follows. See \cite{BH08} for more examples: 
The Cauchy distribution and autocorrelated data from U(1) lattice 
gauge theory.
\begin{figure}[ht] \begin{center}
\includegraphics[scale=0.8]{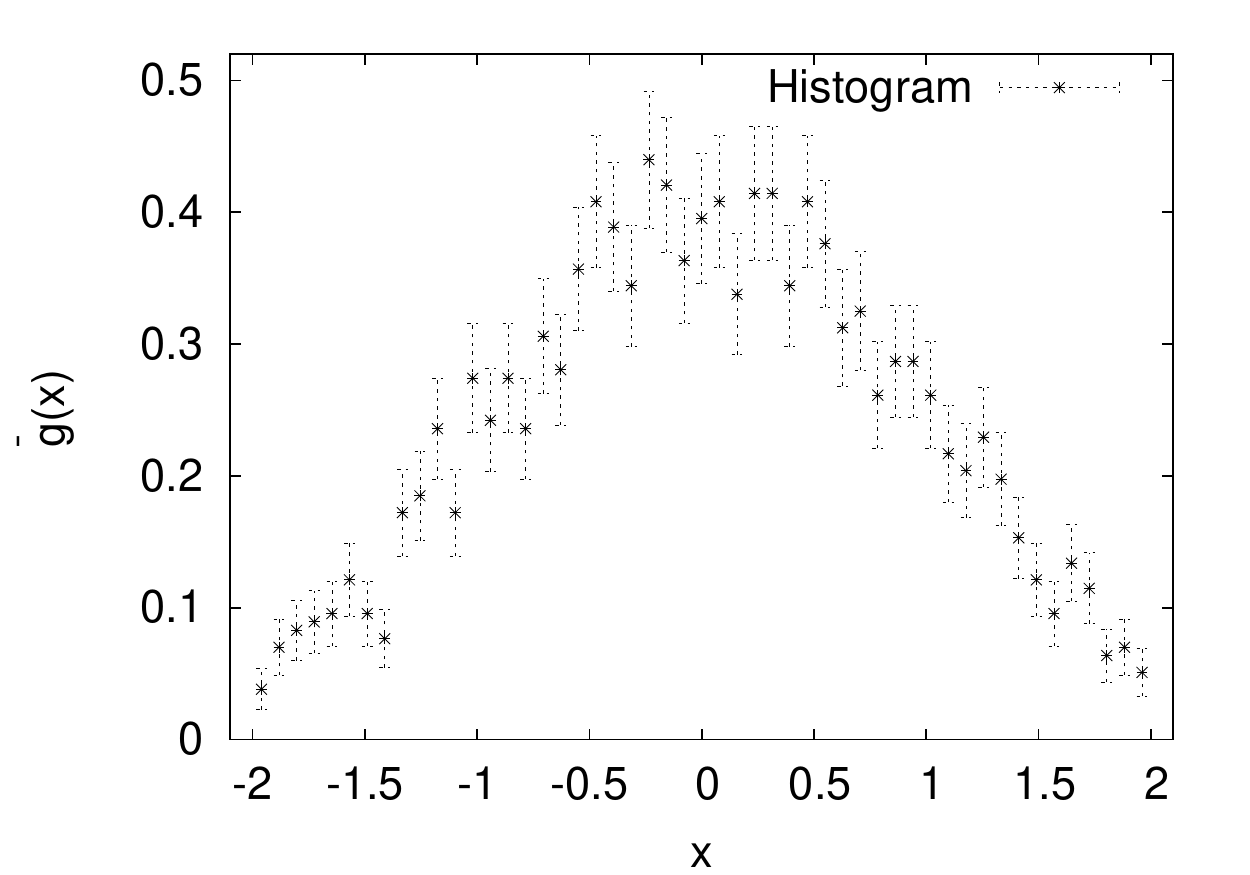}
\caption{\label{fig_GHist} Histogram of 51 bins for 2$\,$000 
random numbers generated according to the Gaussian distribution.}
\end{center} \end{figure}
The histogram for the Gaussian distribution is shown in 
Fig.~\ref{fig_GHist} (the error bars follow from the variance 
$p\,(1-p)$ of the bimodal distribution with $p=h(i)/n$).
\begin{figure}[ht] \begin{center}
\includegraphics[scale=0.8]{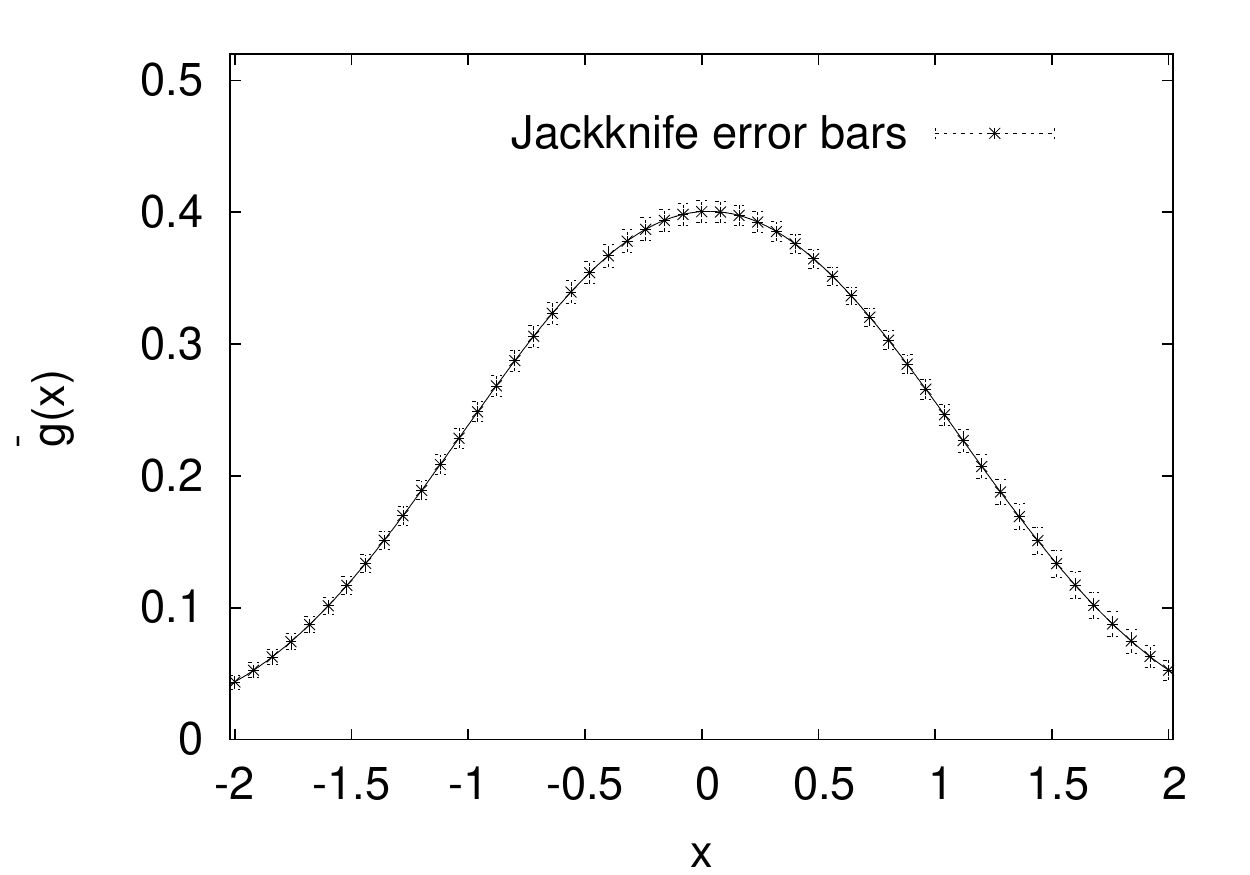}
\caption{\label{fig_GJ} Estimate $\overline{g}(x)$ 
for the data of the previous figure.}
\end{center} \end{figure}
Fig.~\ref{fig_GJ} gives our estimate $\overline{g}(x)$ of the PD 
obtained from the same data with the described method. We used 
$a=x_{\pi_1}$ and $b=x_{\pi_n}$. $Q=0.97$ was reached with $m=4$ 
($Q=0.056$ with $m=3$). Twenty jackknife blocks were used to 
calculate the error bars.

\section{Summary and Conclusions} \label{sec_sum}

Based on Fourier expansion and Kolmogorov tests, we introduced a 
method for constructing continuous probability density functions from 
data. We did not develop a statistically rigorous approach. We address
physicists and others, who do not hesitate to use whatever works.

Our results were obtained with a straight line as initial approximation 
for the CDF. There is certainly space for improvement at the price of 
giving up some of the simplicity. 
With our $Q_{\rm cut}=1/2$ rule, we are slightly overexpanding the 
Fourier expansion. In the average $Q$ should be 1/2, but all our values 
are $Q \ge 1/2$. That gives some flexibility to lower $Q_{\rm cut}$ 
when the $m$ of the Fourier expansion appears to be too large. 

There are many open questions.
Given the initial approximation, we construct a smooth Fourier 
expansion of the remainder, that is consistent with the data, using 
the ordering in which the long wave lengths modes come first. 
Obviously, the result of this procedure is not the only analytical 
function, which is consistent with the data. 
Which ordering of the serious expansion or other complete function 
system gives the smoothest approximation (smallest number of terms) 
consistent with the data?
Do systems of monotonically increasing functions exist, which are
complete for the expansion of monotonically increasing functions?

{\it Kernel density estimates} \cite{J09,Wiki} are in spirit similar 
(but by no means identical) to our method. A comparison remains to 
be carried out.

\section*{Acknowledgments}

This work was in part supported by the US Department of Energy 
under contract DE-FG02-97ER41022.

\end{document}